\documentclass[iop]{emulateapj}
\usepackage{aas_macros}

\usepackage{color}

\usepackage[dvipsnames,svgnames]{xcolor} 
\usepackage{natbib} 
\usepackage{hyperref}
\hypersetup{    
  colorlinks      = true,
  linkcolor       = {black},
  linkbordercolor = {white},
  citecolor       = {black},
  citebordercolor = {white},
  urlcolor        = {blue},
  urlbordercolor  = {white},
}

\usepackage{soul} 
\usepackage{amsmath}
\usepackage{xspace}
\usepackage{lineno}

\newcommand{\code}[1]{\texttt{#1}\xspace}
\newcommand{\degree}{\ensuremath{{}^{\circ}}\xspace}
\newcommand{\unit}[1]{\ensuremath{\mathrm{\,#1}}\xspace}
\newcommand{\feh}{\unit{[Fe/H]}}
\newcommand{\logg}{\unit{log}\ \ensuremath{g}}
\newcommand{\kms}{\unit{km~s^{-1}}}
\mathchardef\mhyphen="2D
\graphicspath{{./fig/}}

\shorttitle{Halo Substructure with Giants}
\shortauthors{Li et al.}

\begin{document}

\title{Exploring Halo Substructure with Giant Stars. XV. Discovery of a Connection between the Monoceros Ring and the Triangulum-Andromeda Overdensity?\altaffilmark{*}\altaffilmark{\dag}\altaffilmark{\ddag}}

\altaffiltext{*}{This paper includes data taken at The McDonald Observatory of The University of Texas at Austin.  }
\altaffiltext{\dag}{This work is based on observations obtained at the MDM Observatory, operated by Dartmouth College, Columbia University, Ohio State University, Ohio University, and the University of Michigan.}
\altaffiltext{\ddag}{Based on observations at Kitt Peak National Observatory, National Optical Astronomy Observatory, which is operated by the Association of Universities for Research in Astronomy (AURA) under cooperative agreement with the National Science Foundation.}


\author{
Ting S. Li\altaffilmark{1,2,3}, 
Allyson A. Sheffield\altaffilmark{4}, 
Kathryn V. Johnston\altaffilmark{5},  
Jennifer L. Marshall\altaffilmark{2,3}, 
Steven R. Majewski\altaffilmark{6},
Adrian M. Price-Whelan\altaffilmark{7},
Guillermo J. Damke\altaffilmark{6,8},
Rachael L. Beaton\altaffilmark{9}, 
Edouard J. Bernard\altaffilmark{10},
Whitney Richardson\altaffilmark{6}, 
Sanjib Sharma\altaffilmark{11},
Branimir Sesar\altaffilmark{12}
}
\altaffiltext{1}{Fermi National Accelerator Laboratory, P. O. Box 500, Batavia, IL 60510, USA}\email{tingli@fnal.gov}
\altaffiltext{2}{Department of Physics \& Astronomy, Texas A \& M University, College Station, TX 77840, USA}
\altaffiltext{3}{George P. and Cynthia Woods Mitchell Institute for Fundamental Physics and Astronomy, College Station, TX, 77843-4242, USA}
\altaffiltext{4}{City University of New York, LaGuardia Community College, Department of Natural Sciences, 31-10 Thomson Ave., Long Island City, NY 11101, USA}
\altaffiltext{5}{Department of Astronomy, Columbia University, Mail Code 5246, New York, NY 10027, USA}
\altaffiltext{6}{Department of Astronomy, University of Virginia, P.O. Box 400325, Charlottesville, VA 22904, USA}
\altaffiltext{7}{Department of Astrophysical Sciences, Princeton University, Princeton, NJ 08544, USA}
\altaffiltext{8}{Departamento de F\'{i}sica y Astronom\'{i}a, Facultad de Ciencias, Universidad de La Serena, Cisternas 1200, La Serena, Chile}
\altaffiltext{9}{The Carnegie Observatories, 813 Santa Barbara Street, Pasadena, CA 91101, USA}
\altaffiltext{10}{Universit\'e C\^ote d'Azur, OCA, CNRS, Lagrange, France}
\altaffiltext{11}{Sydney Institute for Astronomy, School of Physics, University of Sydney, NSW 2006, Australia}
\altaffiltext{12}{Max Planck Institute for Astronomy, K{\"o}nigstuhl 17, D-69117 Heidelberg, Germany;}

\begin{abstract}
Thanks to modern sky surveys, over twenty stellar streams and overdensity structures have been discovered in the halo of the Milky Way. In this paper, we present an analysis of spectroscopic observations of individual stars from one such structure, ``A13'', first identified as an overdensity using the M giant catalog from the Two Micron All-Sky Survey. Our spectroscopic observations show that stars identified with A13 have a velocity dispersion of $\lesssim$ 40~\kms, implying that it is a genuine coherent structure rather than a chance super-position of random halo stars. From its position on the sky, distance ($\sim$15~kpc heliocentric), and kinematical properties, A13 is likely to be an extension of another low Galactic latitude substructure -- the Galactic Anticenter Stellar Structure (also known as the Monoceros Ring) -- towards smaller Galactic longitude and farther distance. Furthermore, the kinematics of A13 also connect it with another structure in the southern Galactic hemisphere -- the Triangulum-Andromeda overdensity. We discuss these three connected structures within the context of a previously proposed scenario that one or all of these features originate from the disk of the Milky Way.

\end{abstract}


\section{Introduction}\label{sec:intro}
Over the last two decades, large-area digital sky surveys such as the Two Micron All Sky Survey~\citep[hereafter 2MASS;][]{Skrutskie2006} and the Sloan Digital Sky Survey~\citep[hereafter SDSS;][]{York2000}, have provided deep and global photometric catalogs of stars in the Milky Way. A variety of substructures in the Galactic halo have been revealed as a result of mapping the Milky Way with these modern surveys using various stellar tracers. The most prominent structures are the tidal tails of the Sagittarius dwarf galaxy~\citep{Ibata1994, Majewski2003}, which provide dramatic evidence that the Milky Way is still being shaped by the infall and merging of neighboring smaller galaxies. These observational results have lent strong support to the hierarchical picture of galaxy formation under the $\Lambda$CDM model~\citep{Bullock2001, Johnston2008, Helmi2011}. 


While the discovery of local overdensities in stellar surveys by visual inspection has proven successful, the scale and sophistication of the data provided by current and future sky surveys motivate an exploration of methods that can instead objectively and automatically identify structures \citep{Sharma2011}. 
\citet{2011ApJ...728..106S} developed a density-based hierarchical group finding algorithm \code{Enlink} to identify stellar halo substructures and applied it to a catalog of M giant stars selected from 2MASS \citep[][hereafter S10]{Sharma2010}. This algorithm uncovered 16 candidate substructures in the Milky Way halo, six of which had not been previously identified.

This paper presents the moderate-resolution spectroscopic analysis of stars in one of the substructures reported by S10, namely the A13 candidate. The goal of this study is to determine, using their kinematical and chemical properties, whether the M giants in A13 are genuinely associated with each other  --- an important test of the performance of the group finding algorithm for finding substructure in the Milky Way halo using 2MASS photometry. 

This work also aims to explore associations between A13 and other known substructures. Its position on the sky -- close to the Galactic anticenter with Galactic latitude $b\sim30\degree$ -- is suggestive of a possible connection between this and two low Galactic latitude structures -- the Monoceros Ring (Mon) and the Triangulum-Andromeda overdensity (TriAnd).

Mon was first discovered by~\citet{Newberg2002} using main-sequence turnoff stars selected from SDSS. Mon appears to be a low Galactic latitude ring-like structure near the Galactic anticenter. It was independently mapped using M giants selected from the 2MASS catalog by~\citet{Crane2003} and \citet{Rocha2003} (who refer to it as the Galactic Anticenter Stellar Structure -- GASS -- in their work). \cite{Crane2003} report the properties they derive from spectra of a group of 53 M giants in the direction of the overdensity they identify as GASS/Mon. They find a velocity dispersion of $\sigma_v \sim 20 \kms$ and a mean metallicity of $\feh = -0.4\pm 0.3$ and derive heliocentric distances of these stars in the range $d \sim 10-12$~kpc. 

Soon after, the Triangulum-Andromeda cloud (TriAnd) was discovered as a ``cloud-like" spatial overdensity by~\citet{Rocha2004} using M giants from the 2MASS catalog, spanning the range $100\degree < l < 160\degree$ and $-50\degree < b < -15\degree$.
\citet{Majewski2004b} also identified a main-sequence turnoff in the foreground of a M31 halo survey that is consistent with the M giant population identified by~\citet{Rocha2004}.
\citet{Martin2007} subsequently detected a second main-sequence in the region of TriAnd -- referred to as TriAnd2 -- using the imaging data from the MegaCam Survey.
\citet{Sheffield2014} conducted an expanded survey of M giants in the TriAnd region using the 2MASS catalog, and found a Red Giant Branch (RGB) sequence at brighter apparent magnitude in addition to the fainter one discovered by~\citet{Rocha2004}. The brighter sequence (refered to as TriAnd1) was assessed to be younger (6-10 Gyr) and closer (distance of $\sim$15-21~kpc), while the fainter sequence (TriAnd2) was found to be older (10-12 Gyr) and farther ($\sim$24-32~kpc). The velocity dispersion for each sequence is $\sigma_v \sim 25 \kms$. 

We note that several low Galactic latitude structures were also discovered near the Milky Way anti-center which may be associated with Monoceros Ring, such as the Anticenter Stream~\citep{Grillmair2006b, Belokurov2007} and the Eastern Banded Structure~\citep{Grillmair2006b, Grillmair2011}. We will not include these structures in this paper as these were detected as overdensities using F turnoff stars in SDSS data, while our study focuses on the structures using M giant stars as the tracer.

\begin{figure}[th!]
\centering
\epsscale{1.2}
\plotone{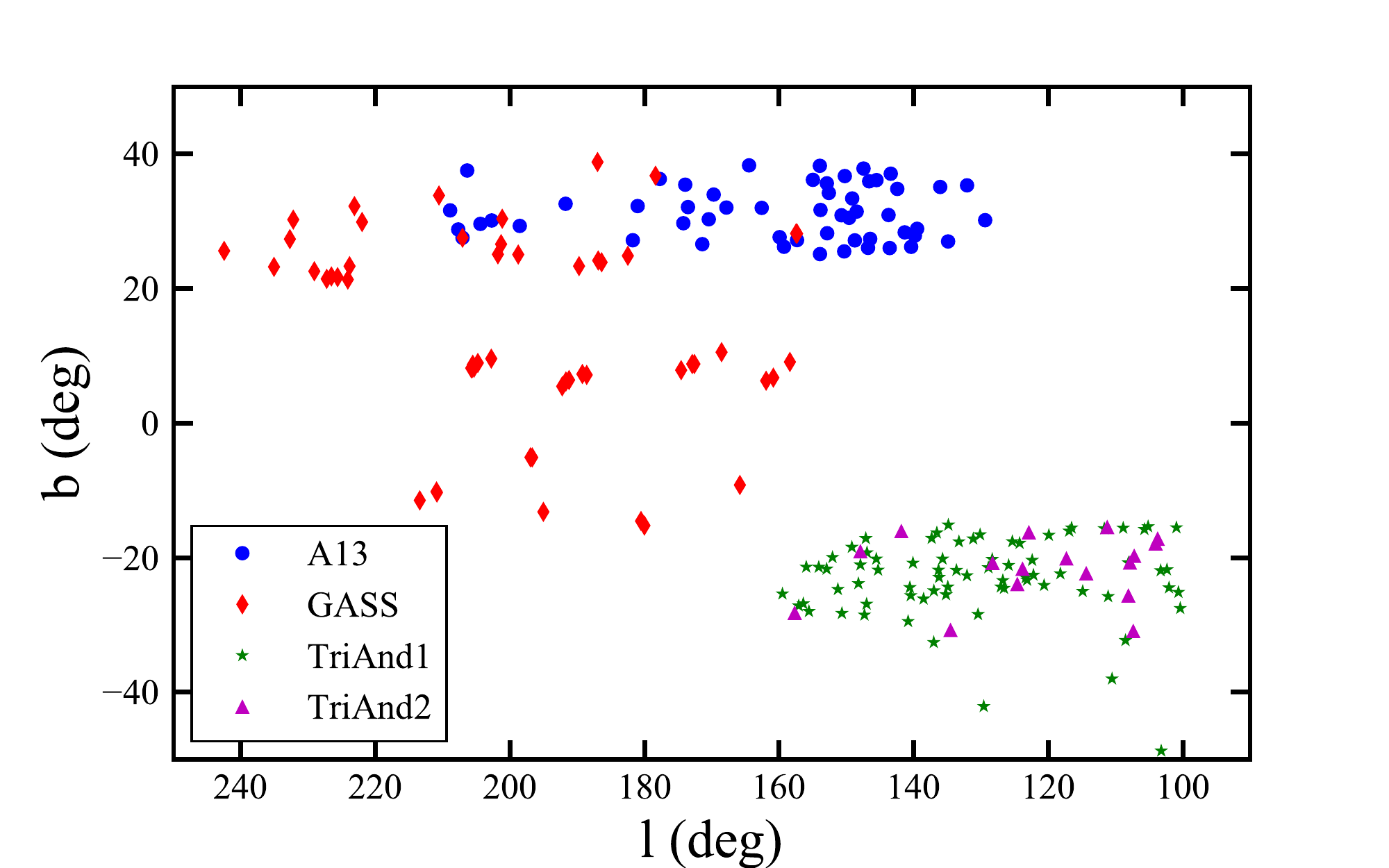}
\caption{Spatial distribution (in Galactic coordinates) of the M giants in the known structures GASS (red diamonds), TriAnd1 (green stars), TriAnd2 (purple triangles), and in the new candidate substructure A13 (blue circles).}
\label{fig:spatial}
\end{figure}

Figure~\ref{fig:spatial} summarizes the distributions of M giant stars in all these structures --- GASS/Mon, TriAnd1, TriAnd2 and A13 --- on the sky. The GASS/Mon sample is from~\citet{Crane2003}, the TriAnd1 and TriAnd2 samples are from~\citet{Sheffield2014}. The A13 sample is from~\citet{Sharma2010} and the results of our spectroscopic follow-up of these A13 stars are presented in this paper.
Our own study is intended to shed light on how these overdensities were formed and may be related to each other by adding position, kinematics and metallicities for the stars in A13 and allowing a direct comparison between all the structures using a single tracer. 
The origins of both GASS/Mon and TriAnd are still under debate. While many studies argue that GASS/Mon and TriAnd could be the remnants of past accretion events~\citep[see, e.g.,][]{Crane2003, Martin2004, Penarrubia2005, Sollima2011, Slater2014, Sheffield2014}, more recent works suggest the possibility that GASS/Mon and TriAnd may be the result of a strong oscillation in the outer disk that throws disk stars to large scale heights~\citep[see, e.g.,][]{Xu2015, PriceWhelan2015}.
A clear picture of the origins of these stellar structures may also have more general implications for how the Milky Way formed. In particular, the extent to which the stellar halo has grown from stars accreted from other systems relative to stars formed in our own Galaxy.

This paper is organized as follows. Section \ref{sec:data} describes our sample, the observations and data reductions. In Section \ref{sec:results}, we present the properties of our target stars derived from the spectra, including their kinematics, metallicities and distances. In Section \ref{sec:discussion} we present a discussion of the possible connection of A13 to GASS/Mon and TriAnd. Section \ref{sec:conclusion} summarizes the results.

\section{Observation and Data Reductions}\label{sec:data}
\subsection{Target selection}\label{sec:sample}
The spectroscopic targets for this study were selected from the 2MASS catalog~\citep{Skrutskie2006} with dereddening applied star-by-star using the~\citet{Schlegel1998} extinction maps. More detailed selection criteria are provided in S10. We give a brief description here. 

M giants begin to separate from M dwarfs in the near-infrared $(J-H, J-K_{s})$ color--color diagram~\citep{Bessell1988}, so that M giant candidates can be efficiently selected using only near-infrared photometry.
In S10, the stars were selected to be M giants in the Galactic halo by applying a selection criteria similar to those used by~\citet{Majewski2003} to identify the tidal tails of the Sagittarius dwarf galaxy.
While~\citet{Majewski2003} restricted their sample to stars with $ (J-K_S)_0 > 0.85 $, S10 selected stars with $ (J-K_S)_0 > 0.97 $ and $K_{S,0} > 10$ (labeled with subscript 0 for dereddened 2MASS photometry, hereafter). 
The first condition in S10 (i.e., $ (J-K_S)_0 > 0.97$) minimized the contamination by disk M dwarfs by restricting the sample to a redder population and the second criterion (i.e., $K_{S,0} > 10$) is chosen to probe deeper into the Galactic halo.
Furthermore, regions of high extinction (typically at low Galactic latitude) were masked from the analysis.
The resulting M giant candidates were further subjected to a group-finding algorithm \code{Enlink}~\citep{Sharma2009} to locate overdense regions, and A13 was one of the overdensities revealed by \code{Enlink} in S10.

Based on the analysis of S10, A13 contains 54 candidate M giant stars, spanning $130\degree<l<210\degree$ and $25\degree<b<40\degree$, lying just north of the edge of one of the rectangular masks ($b\sim25\degree$) for high extinction regions (see Figure 1 or Figure 7 in S10 for the locations of the masks). The range in brightness of the stars in A13 is $10<K_{S,0}<11.3$ and the estimated distance is 23$\pm$11~kpc. The properties of these 54 program stars are listed in Table \ref{table:stars}.
In this study, we have targeted all 54 stars from S10 spectroscopically to further understand the nature of A13.

\begin{deluxetable*}{cccccccr}
\setlength{\tabcolsep}{0.06in}
\tabletypesize{\scriptsize}
\tablecolumns{12}
\tablewidth{0pc}
\tablecaption{
Properties of the Program Stars
\label{table:stars}
}
\tablehead{    
\colhead{ID} & \colhead{RA} & \colhead{Dec} & \colhead{$l$} & \colhead{$b$} &\colhead{$K_{S,0}$} & \colhead{($J-K_S$)$_0$} & \colhead{$v_{hel}$(\kms)}
}
\startdata

A13-01 & 08:22:35.1 & 16:57:51 & 207.06 & 27.56 & 10.07 & 1.04 & 100.2\\
A13-02 & 08:40:52.1 & 17:00:08 & 208.90 & 31.63 & 10.53 & 0.99 & $-$38.0\\
A13-03 & 08:25:16.8 & 31:06:39 & 191.74 & 32.61 & 10.35 & 0.98 & $-$45.7\\
A13-04 & 07:49:24.3 & 38:11:16 & 181.75 & 27.19 & 10.20 & 1.01 & $-$33.3\\
A13-05 & 08:14:17.2 & 39:52:39 & 181.03 & 32.27 & 10.30 & 0.98 &  37.7\\
A13-06 & 08:27:04.4 & 19:52:58 & 204.40 & 29.62 & 10.76 & 0.98 & $-$98.2\\
A13-07 & 08:26:58.0 & 21:25:59 & 202.72 & 30.13 & 10.15 & 1.05 & $-$27.7\\
A13-08 & 08:27:35.9 & 46:05:21 & 173.96 & 35.45 & 10.84 & 0.99 & 122.3\\
A13-09 & 07:55:14.5 & 45:10:21 & 174.24 & 29.72 & 10.55 & 0.99 &  14.2\\
A13-10 & 09:01:20.1 & 21:00:55 & 206.37 & 37.56 & 10.71 & 0.97 & $-$33.9\\
A13-11 & 08:04:53.2 & 50:54:55 & 167.84 & 32.05 & 10.13 & 1.00 & $-$48.5\\
A13-12 & 07:18:48.4 & 57:37:29 & 159.28 & 26.22 & 10.25 & 1.01 & $-$64.5\\
A13-13 & 07:58:35.2 & 71:26:37 & 143.74 & 30.94 & 10.80 & 0.98 & $-$54.5\\
A13-14 & 07:59:57.8 & 67:21:15 & 148.48 & 31.45 & 10.30 & 0.98 & $-$84.3\\
A13-15 & 08:23:01.7 & 63:36:56 & 152.61 & 34.22 & 10.38 & 1.04 & $-$62.3\\
A13-16 & 08:35:19.3 & 63:09:13 & 152.91 & 35.65 & 10.35 & 1.07 & $-$80.5\\
A13-17 & 08:37:16.8 & 61:24:39 & 154.98 & 36.17 & 10.06 & 1.11 & $-$16.3\\
A13-18 & 08:02:42.9 & 55:21:17 & 162.59 & 32.00 & 10.36 & 1.04 &  8.7\\
A13-19 & 07:29:03.3 & 63:32:15 & 152.86 & 28.23 & 10.60 & 1.00 & $-$14.2\\
A13-20 & 07:01:29.8 & 65:26:34 & 150.33 & 25.53 & 10.20 & 1.03 & $-$53.1\\
A13-21 & 07:03:17.7 & 62:10:20 & 153.93 & 25.13 & 10.01 & 0.98 & $-$31.8\\
A13-22 & 08:08:19.0 & 46:06:54 & 173.55 & 32.12 & 10.11 & 1.05 &  9.3\\
A13-23 & 08:19:09.0 & 66:35:07 & 149.14 & 33.40 & 10.39 & 1.01 & $-$86.6\\
A13-24 & 08:50:27.1 & 83:34:26 & 129.38 & 30.17 & 10.41 & 0.98 & $-$161.6\\
A13-25 & 06:53:46.3 & 74:23:42 & 140.38 & 26.21 & 10.76 & 1.00 & $-$74.0\\
A13-26 & 06:59:51.4 & 79:21:22 & 134.89 & 27.01 & 10.87 & 0.98 & $-$171.8\\
A13-27 & 08:45:58.5 & 53:38:45 & 164.47 & 38.32 & 10.29 & 1.07 & $-$54.3\\
A13-28 & 07:16:00.8 & 69:06:09 & 146.47 & 27.39 & 11.03 & 1.01 & $-$45.2\\
A13-29 & 07:15:42.5 & 67:03:59 & 148.75 & 27.17 & 10.13 & 0.98 & $-$19.7\\
A13-30 & 06:56:33.8 & 71:31:34 & 143.57 & 26.03 & 10.05 & 1.01 & $-$220.1\\
A13-31 & 08:18:28.7 & 24:35:04 & 198.55 & 29.32 & 10.16 & 1.02 & 71.3\\
A13-32 & 07:34:53.2 & 47:01:07 & 171.42 & 26.61 & 10.98 & 0.98 & $-$19.3\\
A13-33 & 09:26:08.1 & 76:14:13 & 136.06 & 35.11 & 11.03 & 1.01 & $-$63.3\\
A13-34 & 09:06:13.0 & 66:48:04 & 147.46 & 37.85 & 11.28 & 0.98 & $-$7.9\\
A13-35 & 09:11:42.5 & 70:03:57 & 143.41 & 37.08 & 10.62 & 0.99 & $-$33.0\\
A13-36 & 07:49:59.9 & 66:26:33 & 149.60 & 30.52 & 10.34 & 1.09 & $-$65.1\\
A13-37 & 08:55:51.8 & 61:47:33 & 153.95 & 38.26 & 11.01 & 1.02 & $-$26.5\\
A13-38 & 08:17:34.6 & 49:28:19 & 169.73 & 33.98 & 10.18 & 0.97 & 8.3\\
A13-39 & 08:28:13.3 & 16:54:10 & 207.70 & 28.79 & 10.42 & 0.99 & 49.4\\
A13-40 & 08:33:42.2 & 43:06:22 & 177.74 & 36.30 & 10.60 & 1.00 & 15.1\\
A13-41 & 07:59:56.5 & 62:46:26 & 153.85 & 31.69 & 10.39 & 1.04 & $-$82.4\\
A13-42 & 07:01:45.3 & 68:40:26 & 146.79 & 26.06 & 10.50 & 0.99 & $-$49.7\\
A13-43 & 07:24:12.3 & 59:29:12 & 157.34 & 27.21 & 10.17 & 1.09 & $-$7.0\\
A13-44 & 07:29:55.6 & 57:16:14 & 159.93 & 27.65 & 10.34 & 0.97 & $-$11.8\\
A13-45 & 07:55:41.7 & 48:29:54 & 170.46 & 30.31 & 10.63 & 0.99 & 7.9\\
A13-46 & 08:49:13.9 & 65:01:35 & 150.24 & 36.73 & 10.14 & 0.99 & $-$53.0\\
A13-47 & 08:49:36.7 & 71:38:04 & 142.44 & 34.82 & 10.13 & 1.03 & $-$46.5\\
A13-48 & 08:49:41.6 & 68:02:29 & 146.62 & 35.95 & 10.20 & 1.03 & $-$62.9\\
A13-49 & 07:53:27.3 & 65:28:07 & 150.73 & 30.90 & 10.03 & 0.99 & $-$27.0\\
A13-50 & 08:54:29.6 & 68:49:43 & 145.53 & 36.13 & 10.12 & 1.01 & $-$135.9\\
A13-51 & 10:06:27.0 & 78:40:30 & 132.07 & 35.35 & 10.38 & 1.01 & $-$168.4\\
A13-52 & 07:18:30.1 & 75:01:02 & 139.82 & 27.88 & 10.29 & 1.06 & $-$58.5\\
A13-53 & 07:25:52.9 & 73:40:52 & 141.34 & 28.36 & 10.83 & 0.97 & $-$58.1\\
A13-54 & 07:34:21.4 & 75:17:23 & 139.49 & 28.89 & 10.57 & 1.05 & $-$113.1\\
\enddata
\end{deluxetable*}

\begin{deluxetable*}{lccc}
\setlength{\tabcolsep}{0.2in}
\tabletypesize{\scriptsize}
\tablecolumns{4}
\tablewidth{0pc}
\tablecaption{
Summary of Observing Runs
\label{table:obs}
}
\tablehead{
\colhead{UT} & \colhead{Observatory} & \colhead{Telescope} & \colhead{Spectrograph} 
}
\startdata  	
2011 Nov 10     & MDM      & Hiltner 2.4 m      & Modspec \\
2011 Nov 15--20 & KPNO     & 2.1 m              & Goldcam \\
2012 Nov 28--30 & McDonald & Otto Struve 2.1 m  & ES2 \\
2012 Oct 27--29 & MDM      & Hiltner 2.4 m      & Modspec \\
2014 Jan 09--12 & McDonald & Otto Struve 2.1 m  & ES2  \\
\enddata

\end{deluxetable*}

\subsection{Observation and Data Reduction}\label{sec:obs}
Spectra for this work were collected over five observing runs using telescopes at MDM Observatory, Kitt Peak National Observatory (KPNO), and McDonald Observatory. The observing nights, telescopes and instruments are summarized in Table \ref{table:obs}. 
For all five observing runs, biases were taken at the beginning and end of each night to verify that there were no significant variation over time in the zero noise level. Flats were taken every night using a quartz lamp. To ensure accurate radial velocity measurements, calibration arc frames were taken throughout the night at the same sky position as each target, to account for telescope flexure. For both Goldcam and Modspec, XeNeAr lamps were used; for ES2, NeAr lamps were used. 
All three instruments were set up so that they covered the spectral range 8000--8900~\AA~, with a spectral resolution of $\sim$ 4~\AA and a pixel scale of 1.4~\AA/pixel. This spectral range covers both the Na \textsc{i} doublet lines around 8200~\AA, which are used to discriminate foreground M dwarfs~\citep[see, e.g.,][]{Schiavon1997}, and the Ca \textsc{ii} triplet lines (CaT) around 8500--8700~\AA, which are used to derive the radial velocities and metallicities. The target spectra had a mean signal-to-noise ratio (S/N) $\sim$ 25 per pixel around the CaT feature. 
Most spectra taken during the MDM and McDonald runs have S/N $>$ 30 per pixel, while the spectra from the KPNO run have S/N $\sim$ 15 per pixel. Besides the program stars, a handful of radial velocity (RV) standard M giant stars and one telluric standard star were observed every night along with the program stars. 
The RV standards observed are taken from the Astronomical Almanac and have similar spectral types as the program stars.

\begin{figure}[th!]
\centering
\epsscale{1.2}
\plotone{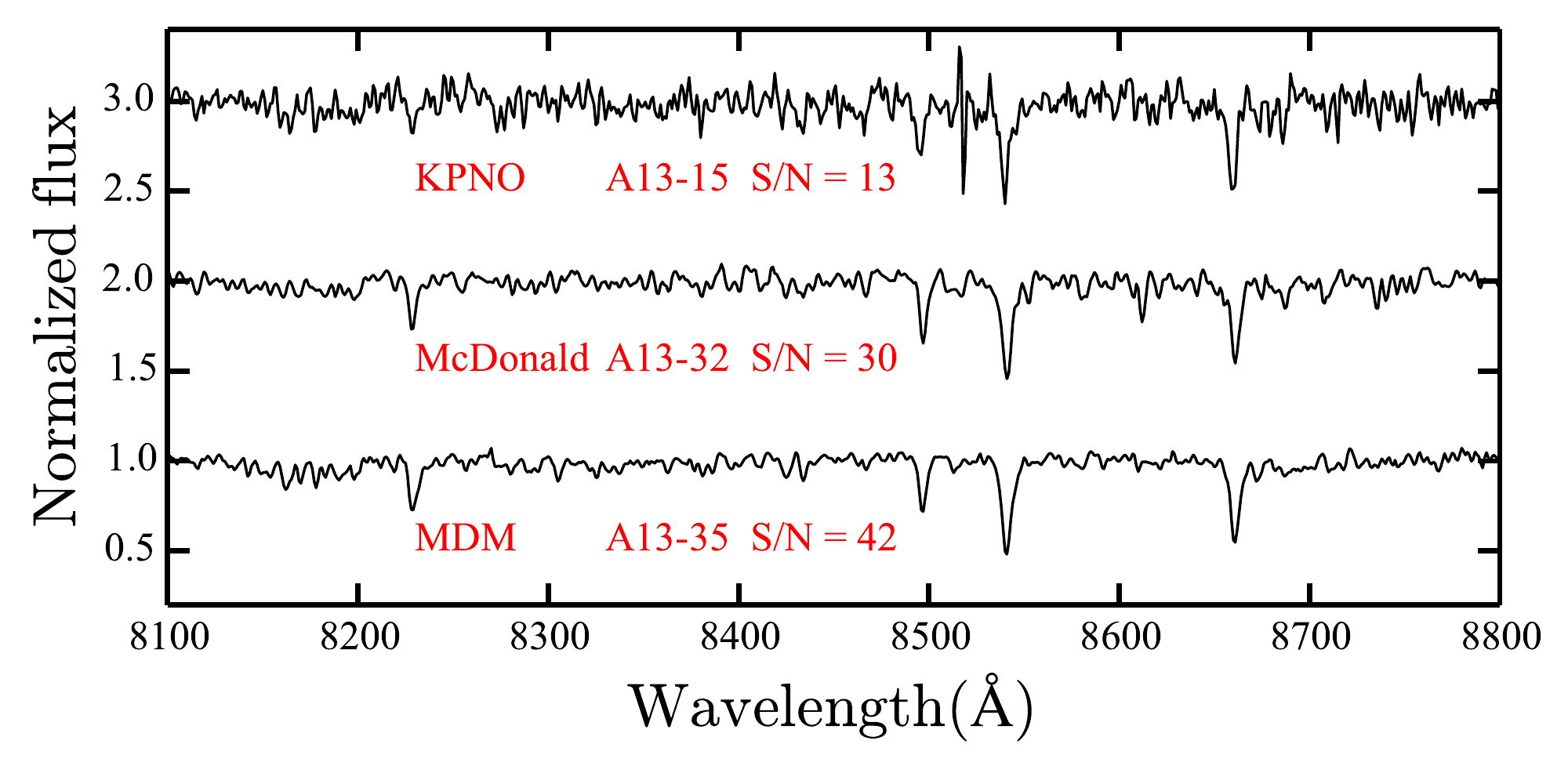}
\caption{Examples of wavelength calibrated spectra of A13 stars obtained with different telescopes/instruments at various S/N. The spectrum of A13-15 represents the lowest S/N and the spectrum of A13-35 represents the highest S/N in our sample.}
\label{fig:samplespec}
\end{figure}

We reduced the data with the standard routines in IRAF. We began by subtracting the bias level using the overscan strip on each frame. A normalized flat was created for each night using the median-combined flats, and the science frames were divided by the normalized flat.  The $apall$ task was used for one-dimensional spectral extraction and the $identify$ task was used for deriving the pixel-to-wavelength calibration. The resulting dispersion solution was applied to the spectra using the $dispcor$ task. Finally, we used $continuum$ task to normalize the continuum of the spectra to one. The wavelength calibrated spectra for several program stars are shown in Figure~\ref{fig:samplespec}. Spectra from three telescopes/instruments show similar resolution and wavelength coverage.

\section{Data Analysis and Results}\label{sec:results}
\subsection{Dwarf/Giant Separation}\label{sec:dwarf}
To check that our sample is purely M giants, with no foreground contamination from M dwarfs, the gravity sensitive Na \textsc{i} (\unit{\lambda\lambda8183,8195}) doublet was analyzed for all of the targets.
We discriminate dwarfs and giants by measuring the equivalent widths (EWs) of the Na \textsc{i} doublet~\citep[see, e.g.,][]{Schiavon1997}.
We first applied a telluric correction to our spectra using the $telluric$ task in IRAF with the telluric standard star to remove the water vapor absorption around 8227~\AA.
We then shifted the spectra to the rest frame using the RV derived in Section \ref{sec:rv}.
Next we numerically integrated the bandpass 8179--8199~\AA~to measure the EW. All 54 candidates have EWs less than 2~\AA, thus confirming that the candidates are all giants.
The lack of dwarf contamination in our sample is consistent with the fact that the candidates are fairly bright ($10<K_{S,0}<11.3$), and the assessment of~\citet{Majewski2003} who state that severe contamination from M dwarfs should only be a concern with $K_{S,0} > 12.5$. This also matches the M dwarf contamination rate seen in the ~\citet{Sheffield2014} sample, where the M dwarf contamination rate is zero at $K_{S,0}<11.5$.

\subsection{Radial Velocities}\label{sec:rv}

The program stars were cross-correlated against the nightly RV standard stars using the $fxcor$ task in IRAF to calculate the relative velocities. The heliocentric velocities of the program stars were derived after the Earth's motion was corrected for using the $rvcorrect$ task.

For every night, the RV standard stars were also cross-correlated against each other to check the level of stability of the instrument. The average velocity precision as determined from cross-correlating the RV standards is $\sim$ 5.3~\kms. Twenty-three out of 54 program stars were observed on multiple runs using different instruments. The standard deviation of velocities from repeated measurements were calculated for each star and the average standard derivation for 23 stars is 5.5~\kms, which is very close to the RV precision derived from cross-correlating the standards for every night. Therefore, we conclude that $\sigma \sim 5.5$~\kms is the velocity precision for this study. The heliocentric radial velocities for all 54 targets are presented in Table \ref{table:stars}. For the stars observed on multiple runs, the averages of velocities are presented.

We then converted the heliocentric radial velocities, $v_{hel}$, to the radial velocities in the Galactic standard of rest (GSR) frame, $v_{GSR}$. This conversion removes the motion of the Sun with respect to the Galactic Center. We adopted the circular orbital velocity of the Milky Way at the Sun's radius as $\Theta_{0} = 236~\kms$~\citep{Bovy2009} and a solar motion of $(U_\odot,~V_\odot,~W_\odot) = (11.1,~12.2,~7.3)~\kms$~\citep{Schonrich2010}.

\begin{figure*}[th!]
\centering
\epsscale{1}
\plotone{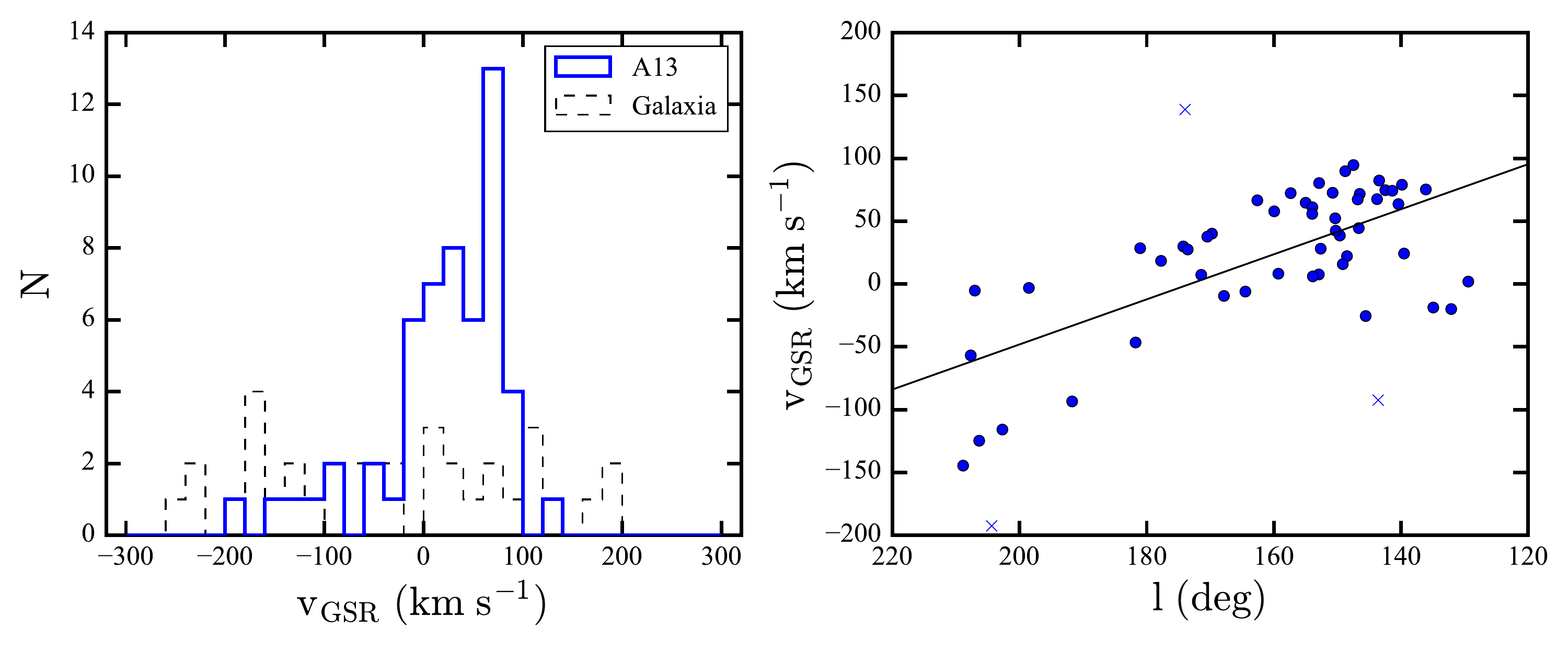}
\caption[Distribution of the radial velocities in the Galactic Standard of Rest frame, $v_{GSR}$, for the stars in the A13 sample.]{Distribution of the radial velocities in the Galactic standard of rest frame, $v_{GSR}$, for the stars in the A13 sample. Left panel: Histogram of $v_{GSR}$ for 54 M giants in A13. A clear peak shows up around 50~\kms. As a comparison, the distribution of $v_{GSR}$ for stars in a mock catalog generated using the Galaxia model~\citep{Sharma2011} with identical photometric and spatial properties as A13 is shown as the black dashed line. Right panel: A scatter plot of $v_{GSR}$ vs. Galactic longitude for 54 M giants in A13. $v_{GSR}$ increases as the Galactic longitude of stars decreases, consistent with prograde rotation. The line shows a linear fit with 2.5-$\sigma$ rejection applied to the data; three stars were removed as outliers, shown as the crosses.}
\label{fig:vgsr_hist}
\end{figure*}

The distribution of $v_{GSR}$ of the 54 stars in A13 is shown in the left panel of Figure~\ref{fig:vgsr_hist}. The stars in A13 show a prominent peak at $v_{GSR}\sim50$~\kms. As a comparison, we also show the velocity distribution of a synthetic sample of Milky Way field stars generated from the Galaxia model~\citep{Sharma2011}. The synthetic sample is selected to be within the same patch of the sky (i.e., $130\degree<l<210\degree$ and $25\degree<b<40\degree$) with the same magnitude and color range (i.e., $0.97 < (J-K_S) < 1.11$ and $10 < K_S < 11.3$) as the A13 sample and is mostly composed of halo stars. It is apparent that the A13 sample has a much smaller velocity dispersion than expected for a random distribution of halo stars as predicted by Galaxia as shown in Figure~\ref{fig:vgsr_hist}.

The RV data also show a velocity gradient in the sense that $v_{GSR}$ is increasing as the Galactic longitude of stars decreases (see right panel of Figure~\ref{fig:vgsr_hist}), consistent with a prograde rotation. We apply a linear fit to $v_{GSR}$ as a function of Galactic longitude, in the process of removing 2.5-$\sigma$ outliers iteratively. Three stars (A13-06, A13-08 and A13-30) are removed as outliers.\footnote{We note that though only three stars are identified as outliers here, most background stars has been discarded when the A13 structure was identified by the group finder in S10. If we apply the same selection criteria for the Galaxia sample to the dereddened 2MASS catalog, we get a total of $\sim$101 stars. The group finder  identified 54 stars as the overdensity structure A13, and considered the remaining $\sim$47 stars as the background. As comparison, we get 37 stars as the background stars from Galaxia, shown as dashed histogram in Figure~\ref{fig:vgsr_hist}. Galaxia predictions are therefore consistent with the observations.}  From a linear fit of the remaining 51 members, we derive the velocity dispersion, the gradient, and the velocity at $l=180\degree$ to be:
\begin{equation}
\sigma_v \lesssim 40 \kms
\end{equation}
\begin{equation}
dv_{GSR}/dl=-1.57 \pm 0.28 \kms \mathrm{deg^{-1}}
\end{equation}
\begin{equation}
v_{GSR|180\degree}=-12.0 \pm 7.9 \kms
\end{equation}

The dispersion of $\sigma_v \lesssim 40 \kms$ is much colder than the velocity dispersion for the kinematics of random halo stars, as shown by the Galaxia model, but hotter than the $v_{GSR}$ dispersion for the Sagittarius tidal stream~\citep[$\sigma_v \sim 10-25$~\kms;][]{Majewski2004a, Gibbons2017} and the Orphan stream~\citep[$\sigma_v \sim 10$~\kms;][]{Newberg2010}. The measured velocity dispersion is likely affected by unidentified outliers, which bias the measured dispersion towards higher values. Still, the velocity dispersion of A13 is closer to that of disk stars.

\subsection{Metallicities}\label{sec:feh}

The Ca \textsc{ii} triplet (CaT) has been historically used to determine the metallicities of the stars in globular and open clusters whose distances are known~\citep[see, e.g.,][]{Armandroff1988, Rutledge1997, Cole2004, Warren2009, Carrera2012}. \citet{Battaglia2008} show that CaT--\feh relations calibrated on globular clusters can also be applied with confidence to RGB stars in composite stellar populations. 
\citet{Carrera2013} derived a relation between the \feh, luminosity, and the EWs of the CaT lines using 500 RGB stars in clusters and 55 metal-poor field stars. To date, many studies have used this calibration relation to derive the metallicities for giant stars, especially in dwarf spheroidal galaxies~\citep[see, e.g.,][]{Hendricks2014, Simon2015}. However, most of the RGB stars used in these calibration works have temperatures higher than our M giant sample. Furthermore, as our sample has no precise distance measurement, the luminosities of the stars are unknown and we cannot use the relation derived from the aforementioned studies.

We therefore developed an empirical method to derive the metallicities of the M giants in the A13 sample from the CaT EWs and assumed a linear relation between the metallicity and the CaT EWs for M giants. This method is similar to that described by ~\citet{Sheffield2014} but with a larger sample of calibration stars. We collected spectra for 22 late spectral type giants with published metallicities in our 2014 McDonald run. To include more calibrators, we extend the range of $(J-Ks)$ color from $(J-Ks) > 0.97$ to $(J-Ks)>0.82$. The information of the 22 red calibrators are listed in Table~\ref{table:feh_cal}. The most metal-poor calibration star, HD 37828, with $(J-Ks)=0.83$, is also the bluest star in the calibration sample. We included this star to get a wider metallicity range even though it is much bluer than our program stars. Including HD\ 37828 should not introduce large additional systematic errors, as the other 21 calibration stars do not show a correlation between \feh and $(J-Ks)$. 

We compute a spectral index for each of the Ca \textsc{ii} triplet lines ($\lambda\lambda$8498, 8542, 8662). The spectral indices are a pseudo EW measured in~\AA, which is defined as: 

\begin{equation}
\mathrm{EW}=\int_{\lambda_1}^{\lambda_2}\left ( 1-\frac{F_l(\lambda)}{F_c(\lambda)} \right )d\lambda
\end{equation}
where $\lambda_1$ and $\lambda_2$ are the edges of bandpasses for the CaT lines,  $F_l(\lambda)$ is the flux of the line and $F_c(\lambda)$ is the continuum flux. The continuum flux is computed as a linear fit using the red and blue continuum bandpasses. We evaluated EWs using the line and continuum bandpass definitions both from~\citet{Du2012} and~\citet{Cenarro2001}. 

\begin{figure*}[th!]
\centering
\epsscale{1.0}
\plottwo{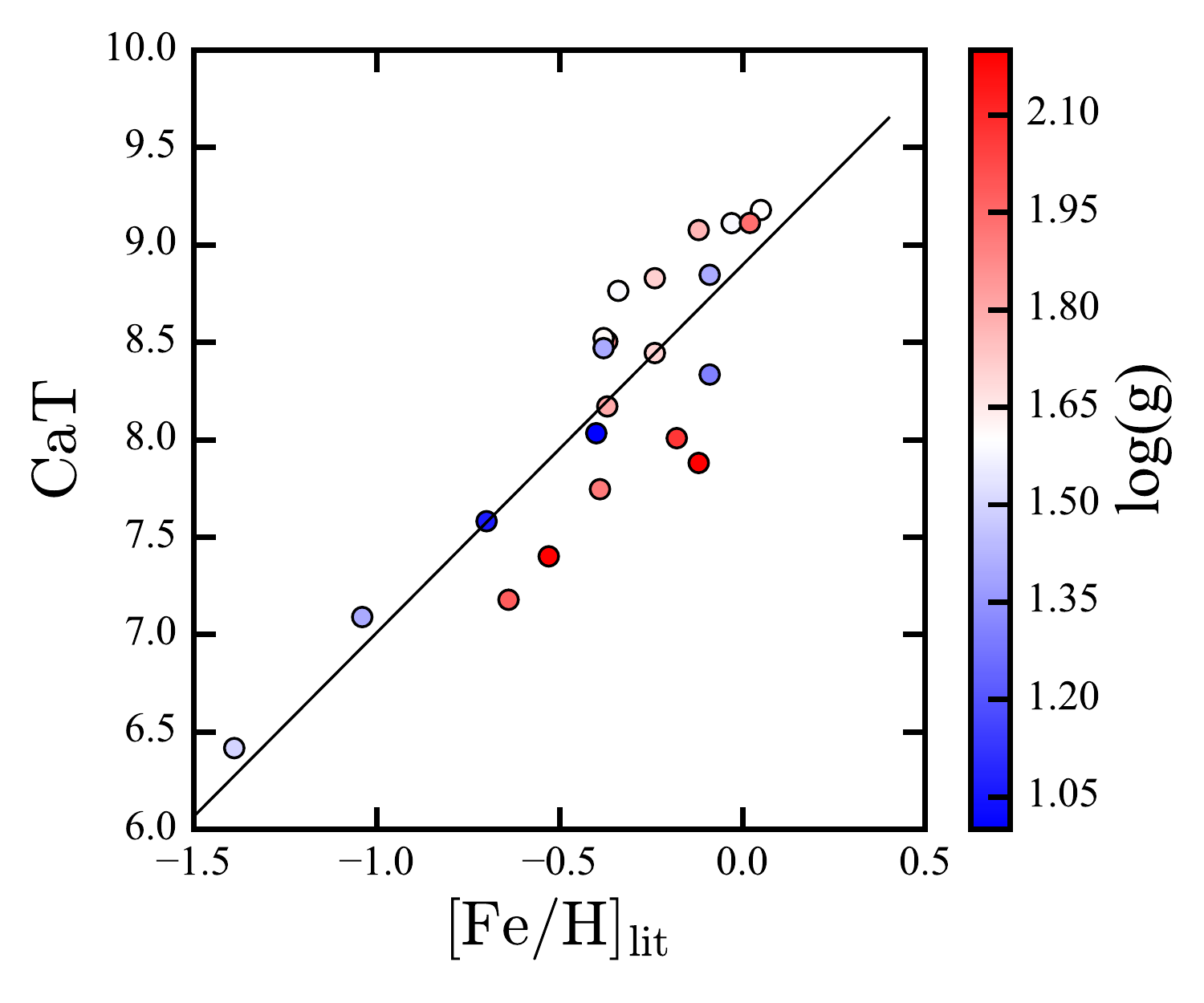}{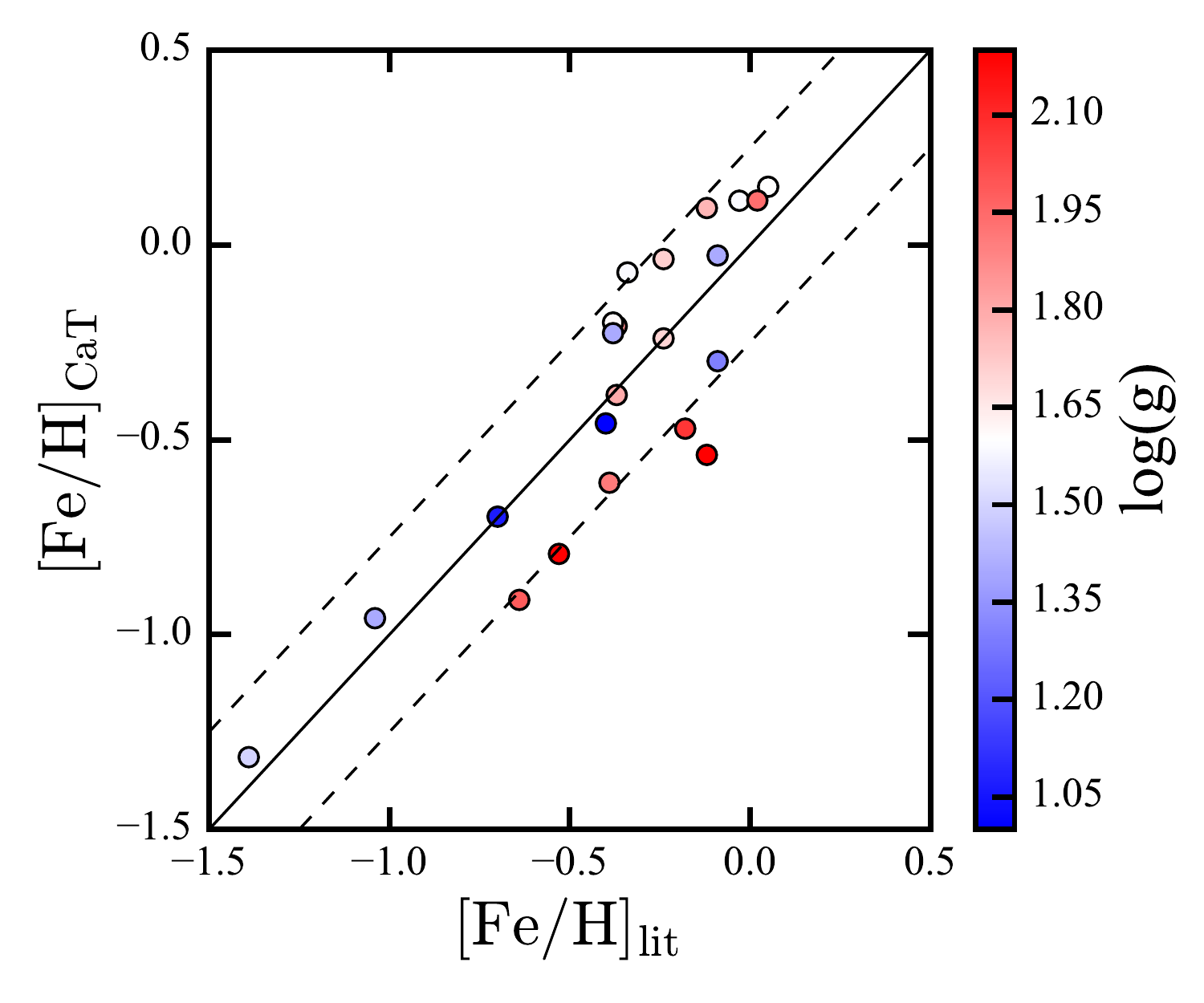}
\caption[Metallicity calibration using 22 giant stars with known metallicities.]{Metallicity calibration using 22 giant stars with known metallicities. Left Panel: The CaT--\feh linear fit for 22 metallicity calibration giant stars, where CaT is the summation of the spectral indices from the near-IR calcium triplet. Right Panel: [Fe/H] derived from the CaT index vs. the published [Fe/H] values. The dashed lines are $\pm$0.25 dex away from the one-to-one (solid) line. For both panels, each symbol is also color-coded with its literature surface gravity, \logg. 
}
\label{fig:feh_cal}
\end{figure*}

We use a simple sum of the three Ca \textsc{ii} spectral indices as the total strength of the Ca triplet CaT, i.e.
\begin{equation}
\mathrm{CaT} = \mathrm{EW_{8498}} + \mathrm{EW_{8542}} + \mathrm{EW_{8662}}
\end{equation}
We then derive a linear empirical relation between the CaT index and the published [Fe/H] for the calibrator stars. The bandpasses defined by~\citet{Du2012} give a smaller standard deviation of the residuals between the published and derived metallicities for the 22 metallicity calibrators. Therefore, we chose to use the bandpasses defined by~\citet{Du2012} for the final metallicity calibration relation. 
Our derived CaT--[Fe/H] relation for the 22 metallicity calibrators
is shown as the solid line in the left panel of Figure \ref{fig:feh_cal}. In
the right panel of Figure~\ref{fig:feh_cal}, the [Fe/H] derived from the CaT index is plotted against the published [Fe/H] values. The dashed lines in the right panel are $\pm$0.25 dex away from the one-to-one (solid) line. The estimated error in the derived metallicities is $\pm$0.25 dex, considering that most of the derived [Fe/H] values for the calibrators fall within 0.25 dex of the published values. In both panels, we also color-code the published $\logg$ value. The stars with higher $\logg$ tend to have smaller CaT at a given metallicity, as expected. As the CaT index is comprised of absorption lines of ionized calcium, the line strength gets smaller for stars with larger surface gravity. The errors caused by the difference in surface gravity are smaller than 0.25 dex for our calibration stars, which have a surface gravity range of $1.0 < \logg < 2.2$. 

We next apply this derived relation to determine metallicities for the 27 out of 51 A13 stars\footnote{Three are rejected as outliers based upon their kinematics.} that have S/N $>$ 25.
The mean \feh derived from the CaT index for the 27 stars in A13 is \feh = $-$0.57 $\pm$ 0.21, 
where $\pm$ 0.21 dex is the standard deviation of the metallicities for the 27 stars. The derived \feh from the 27 A13 stars span from $-1.1$ to $-0.1$, as shown in the left panel of Figure \ref{fig:a13_feh}. It is possible that there are more metal-poor stars belonging to this structure, however the color criterion with $(J-K_S) > 0.97$ biases the sample against metal-poor stars.

As mentioned earlier, to include more calibration stars, we extended the color range of the calibration sample to $(J-K_S)>0.82$, while our program stars have $(J-K_S)>0.97$. As a test of the impact of the color range difference, we computed the CaT -- \feh relation using only the 10 calibrators with $(J-K_S)>0.95$. The derived mean \feh for the 27 program stars changes from $-$0.57 to $-$0.55. Moreover, because the CaT index tends to have a weak correlation with surface gravity, we also compute the CaT -- \feh relation using 11 calibrators with $1.0 < \logg < 1.7$. The derived mean \feh changes from $-$0.57 to $-$0.63. Thus, the systematic errors in the metallicity calibration due to changing the characterization of our calibration sample are less than 0.1 dex.

\begin{deluxetable*}{lcrrrrrrrcc}
\setlength{\tabcolsep}{0.06in}
\tabletypesize{\scriptsize}
\tablecolumns{12}
\tablewidth{0pc}
\tablecaption{
Properties of the Giants for Metallicity Calibration
\label{table:feh_cal}
}
\tablehead{    
\colhead{ID} & \colhead{RA} & \colhead{Dec} & \colhead{$V$} & \colhead{$J$} &\colhead{$K_S$} & \colhead{$T_{\mathrm{eff}}$} & \colhead{$\logg$} & \colhead{\feh} & \colhead{REF\tablenotemark{a}} & \colhead{EW\tablenotemark{b}} 
}
\startdata
HD 5780 & 00:59:23.3 & 00:46:44 & 7.65 & 5.37 & 4.22 & 3848 & 1.07 & $-$0.70 & 1 & 7.58\\
HD 6833 & 01:09:52.3 & 54:44:20 & 6.77 & 5.00 & 4.04 & 4450 & 1.40 & $-$1.04 & 2 & 7.09\\
HD 9138 & 01:30:11.1 & 06:08:38 & 4.84 & 2.52 & 1.66 & 4040 & 1.91 & $-$0.39 & 3 & 7.74\\
HD 13520 & 02:13:13.3 & 44:13:54 & 4.84 & 2.34 & 1.33 & 3970 & 1.70 & $-$0.24 & 3 & 8.44\\
HD 29139 & 04:35:55.2 & 16:30:33 & 0.86 & $-$2.10 & $-$3.04 & 3910 & 1.59 & $-$0.34 & 3 & 8.76\\
HD 30834 & 04:52:38.0 & 36:42:11 & 4.79 & 2.32 & 1.39 & 4130 & 1.86 & $-$0.37 & 3 & 8.50\\
HD 37828 & 05:40:54.6 & $-$11:12:00 & 6.88 & 4.89 & 4.06 & 4430 & 1.50 & $-$1.39 & 4 & 6.41\\
HD 39853 & 05:54:43.6 & $-$11:46:27 & 5.64 & 2.90 & 1.98 & 3994 & 1.00 & $-$0.40 & 1 & 8.03\\
HD 50778 & 06:54:11.4 & $-$12:02:19 & 4.09 & 1.54 & 0.64 & 4000 & 1.80 & $-$0.37 & 3 & 8.17\\
HD 69267 & 08:16:30.9 & 09:11:08 & 3.52 & 1.19 & 0.19 & 4010 & 1.71 & $-$0.24 & 3 & 8.83\\
HD 70272 & 08:22:50.1 & 43:11:17 & 4.26 & 1.26 & 0.38 & 3900 & 1.59 & $-$0.03 & 3 & 9.11\\
HD 81797 & 09:27:35.2 & $-$08:39:31 & 2.00 & $-$0.26 & $-$1.13 & 4120 & 1.77 & $-$0.12 & 3 & 9.07\\
HD 82308 & 09:31:43.2 & 22:58:05 & 4.32 & 1.45 & 0.59 & 3900 & 1.60 & 0.05 & 5 & 9.17\\
HD 90254 & 10:25:15.2 & 08:47:05 & 5.64 & 2.38 & 1.40 & 3706 & 1.40 & $-$0.09 & 6 & 8.85\\
HD 99167 & 11:24:36.6 & $-$10:51:34 & 4.82 & 1.93 & 1.01 & 3930 & 1.61 & $-$0.38 & 3 & 8.52\\
HD 112300 & 12:55:36.2 & 03:23:51 & 3.38 & $-$0.11 & $-$1.19 & 3652 & 1.30 & $-$0.09 & 6 & 8.33\\
HD 115478 & 13:17:15.6 & 13:40:33 & 5.33 & 3.30 & 2.38 & 4240 & 2.21 & $-$0.12 & 3 & 7.88\\
HD 183439 & 19:28:42.3 & 24:39:54 & 4.45 & 1.71 & 0.71 & 3847 & 1.40 & $-$0.38 & 6 & 8.47\\
HD 211073 & 22:13:52.7 & 39:42:54 & 4.51 & 2.26 & 1.29 & 4110 & 1.94 & 0.02 & 3 & 9.11\\
HD 216174 & 22:49:46.3 & 55:54:10 & 5.44 & 3.58 & 2.63 & 4390 & 2.23 & $-$0.53 & 1 & 7.40\\
HD 217459 & 23:00:42.9 & 03:00:42 & 5.85 & 3.91 & 2.94 & 4170 & 2.07 & $-$0.18 & 3 & 8.00\\
HD 220009 & 23:20:20.6 & 05:22:53 & 5.08 & 2.89 & 1.99 & 4435 & 1.98 & $-$0.64 & 7 & 7.17\\
\enddata
\tablenotetext{a}{References: 
(1) \citet{Cenarro2003} (2) \citet{Fulbright2000}, (3) \citet{McWilliam1990}, (4) \citet{Ryan1995}, (5) \citet{Fernandez1990}, (6) \citet{Smith1986}, (7) \citet{Luck2007}}
\tablenotetext{b}{EW is calculated using the bandpasses defined by Du et al. (2012).}
\end{deluxetable*}

\subsection{Heliocentric Distances}\label{sec:distance}
To compute stellar distances, we adopted the  metallicity dependent $M_{K_S}$ -- $(J-K_S)$ relation derived in ~\citet{Sheffield2014}. That is, 

\begin{equation}
M_{K_S} = (3.8+1.3\feh)-8.4(J-K_S).
\end{equation}

\begin{figure*}[th!]
\centering
\epsscale{1.0}
\plottwo{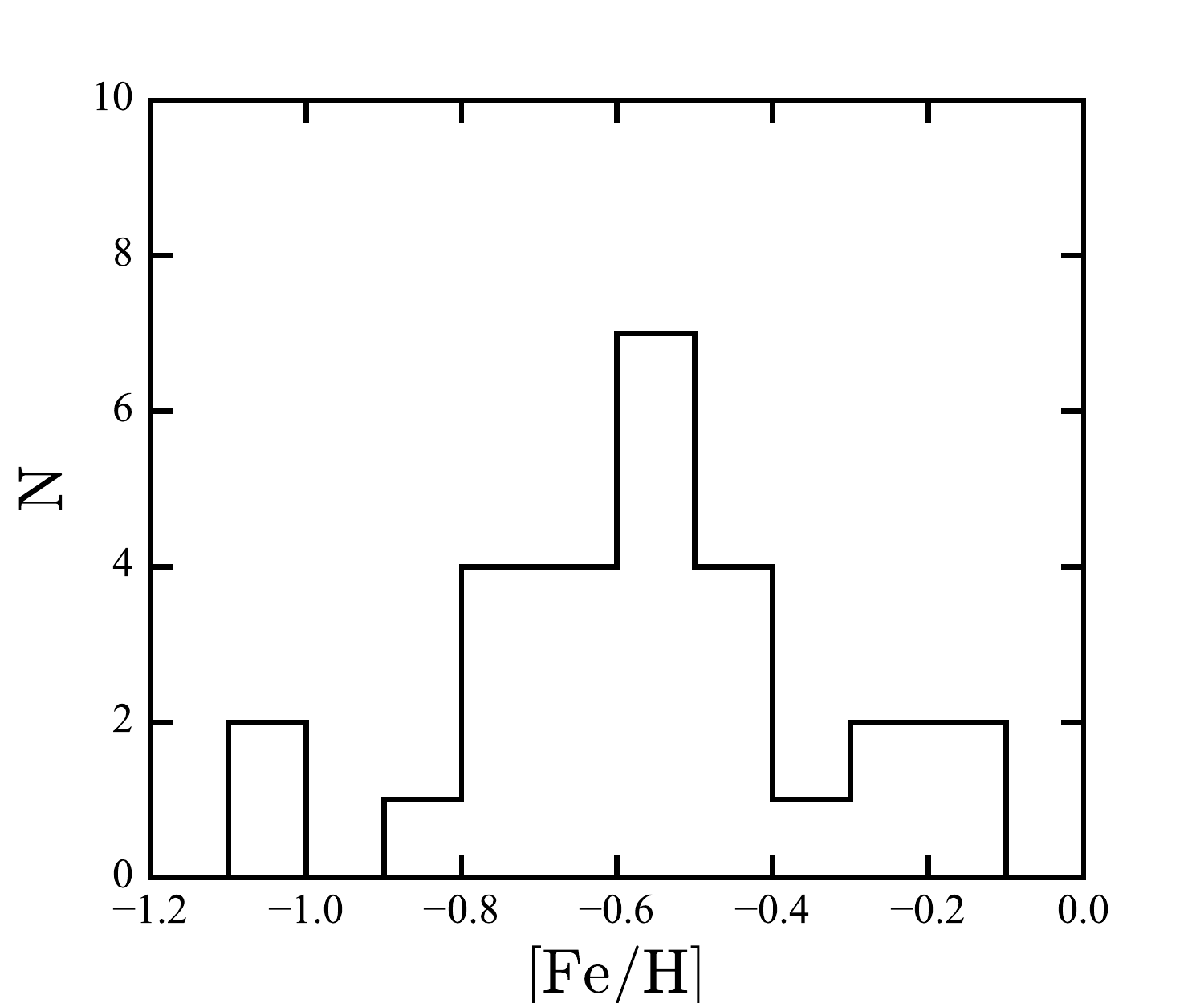}{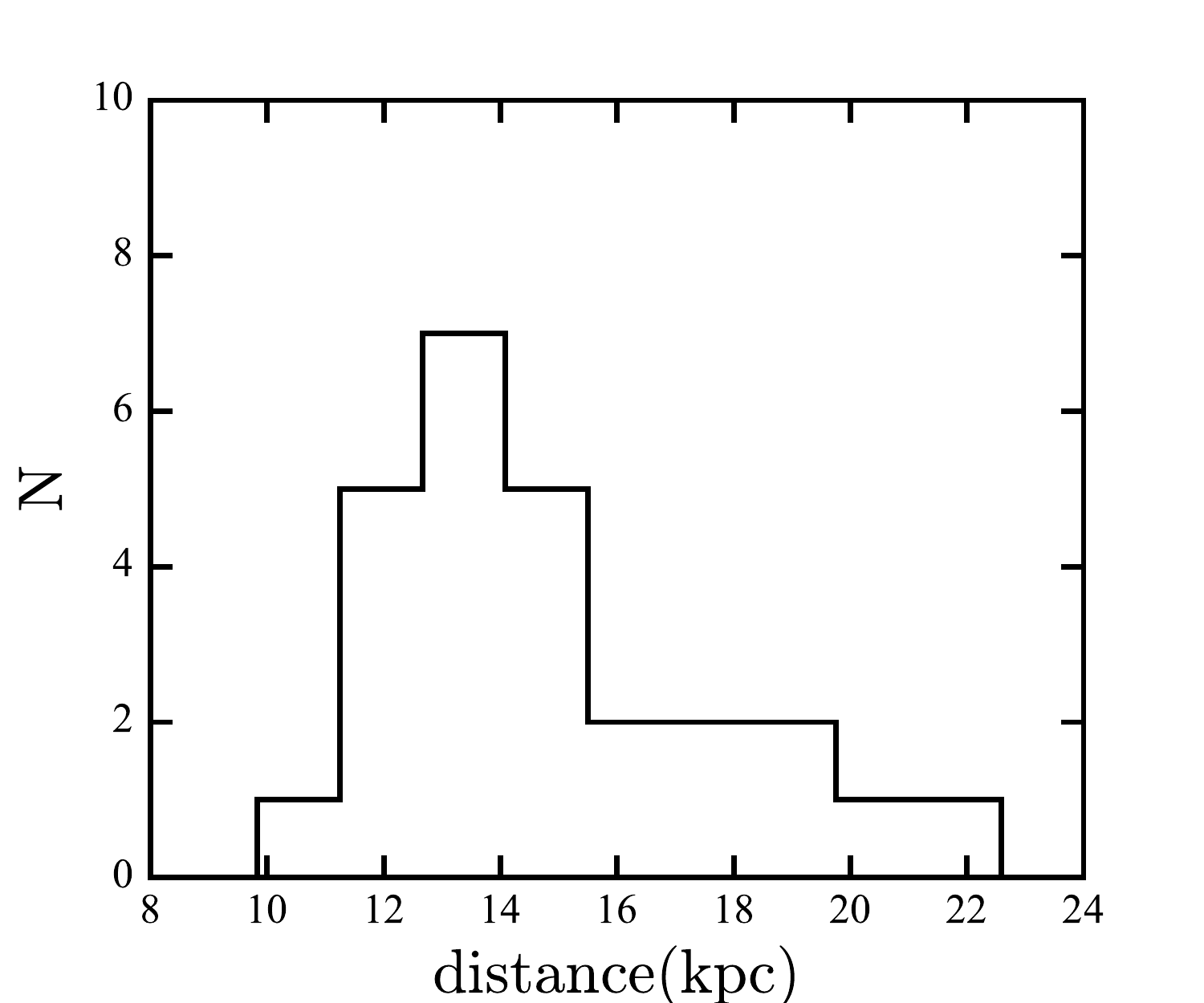}
\caption[Distributions of metallicities and heliocentric distances derived for the A13 sample.]{Distributions of metallicities and heliocentric distances of the A13 sample. Left Panel: Distributions of [Fe/H] derived from summing the EWs for the three calcium triplet lines for 27 A13 stars having S/N $>$ 25. Right Panel: Distributions of heliocentric distances for the same stars.}
\label{fig:a13_feh}
\end{figure*}

For the 27 stars with calculated metallicity, the heliocentric distances were derived individually for each star using the $(J-K_S)$ color and \feh derived previously. As shown in the right panel of Figure \ref{fig:a13_feh}, the heliocentric distances for A13 stars span from 10 to 22 kpc, with a mean of $\sim$15 kpc. S10 estimated the distance for A13 to be 23 $\pm$ 11 kpc based on an assumption of a more metal-poor population.  As an uncertainty of $\pm$0.25 dex in \feh will change $M_{K_S}$ by $\pm$0.32 mag, the uncertainty of the distance for each star is at least 15--20\%.
\textbf{}

\section{Discussion}\label{sec:discussion}
\subsection{Relation to the Galactic Anticenter Stellar Structure}\label{sec:gass}

\begin{figure}[th!]
\centering
\epsscale{1.2}
\plotone{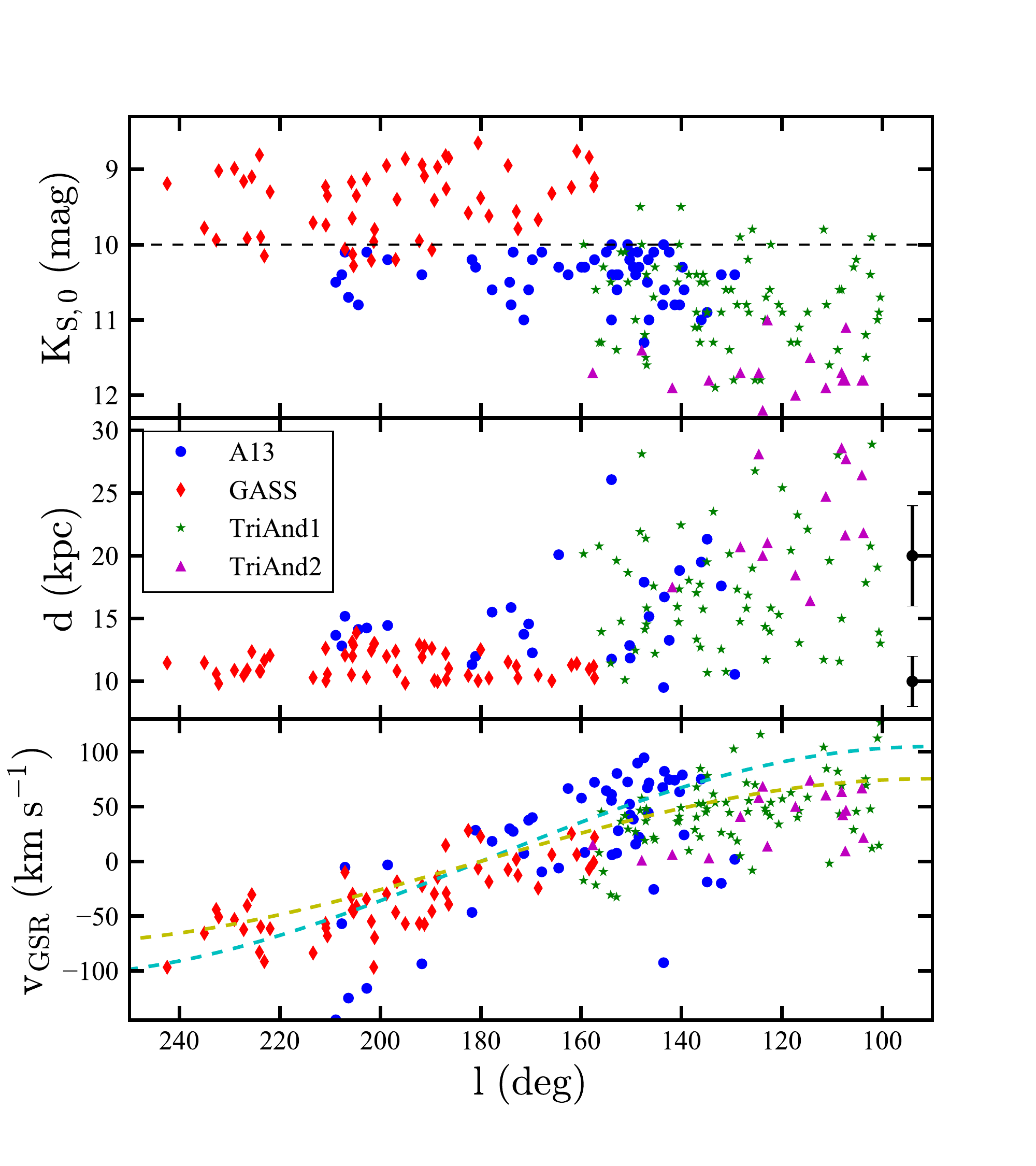}
\caption{A comparison of various properties for A13, GASS, TriAnd1 and TriAnd2. Top Panel:  $K_{S,0}$ as a function of Galactic longitude. Because S10 used a $K_{S,0} > 10$ criterion, most of the A13 giants are fainter than the M giants in the GASS sample of C03. 
Middle Panel: Estimated heliocentric distances of individual M giants using the relation given in Section~\ref{sec:distance} as a function of Galactic longitude. The A13 stars with $l > 180 \degree$ are located at a similar distance to GASS stars while those with $l < 180 \degree$ are located at a farther distance, more like that of TriAnd. 
The black error bars on the right side show the 20\% uncertainty in distances at 10~kpc and 20~kpc.
Bottom Panel: Radial velocities in the GSR frame, $v_{GSR}$, as a function of Galactic longitude. Two dashed curves shows the expected $v_{GSR}$ for an object orbiting circularly at 18~kpc (cyan) and 25~kpc (yellow) with $v_{circ} = 236$~\kms. These are roughly the  Galactocentric distances of GASS and TriAnd, respectively (see Figure~\ref{fig:cartesian}). }
\label{fig:compare}
\end{figure}

We first compare the A13 results with the properties of M giants in GASS/Mon from~\citet[][here after the C03 sample]{Crane2003}. 
As shown in Figure~\ref{fig:spatial}, more than half of the C03 sample are close to the Galactic plane ($|b|<20\degree$), while stars in the A13 sample have $b>25\degree$. This difference in sky positions could be due to selection effects. The C03 sample was not identified by the group-finding algorithm in S10 as part of the A13 group for two reasons: First, about two-thirds of the C03 sample has $b<25\degree$ and therefore they are excluded by the S10 rectangular masks for extinction regions (see Section \ref{sec:sample}). Second, most of the C03 sample has $K_{S,0}<10$ (see the top panel of Figure~\ref{fig:compare}) while S10 made a cut of $K_{S,0}>10$ on their initial sample selection to exclude the nearby stars.

The top panel of Figure~\ref{fig:compare} compares the magnitude distributions of the stars. Because most of the stars in A13 are fainter than GASS, the average heliocentric distance of A13 stars ($d\sim15$~kpc) is farther than that of GASS star ($d\sim11$~kpc), as shown in the middle panel of Figure~\ref{fig:compare}. 

The bottom panel of Figure~\ref{fig:compare} demonstrates that, while the structures may have different locations on the sky, our M giant sample in A13 follows a similar trend in radial velocities as the M giants in GASS/Mon. There is one star observed in both samples, which is the A13-01 in Table~\ref{table:stars}. The observed $v_{hel}$ is $100.2 \pm 5.3$~\kms in our work and $95.6 \pm 2.7$~\kms in C03. The difference is 4.6~\kms, which is within the 1-$\sigma$ joint uncertainty from both measurements. 

Overall, the location on the sky and similar velocity trends suggest that A13 could be a direct extension of GASS/Mon towards smaller Galactic longitude (i.e., $l < 180\degree$) and further heliocentric distance.

\subsection{Relation to the Triangulum-Andromeda Cloud}\label{sec:triand}

We next compare the A13 results with the M giants in TriAnd, presented in ~\citet{Sheffield2014}. 
Member stars in TriAnd1 and TriAnd2 are shown in Figure~\ref{fig:compare}, together with the stars in A13 and GASS.  Though the TriAnd stars are located in the southern Galactic hemisphere, the velocity trends of all of four structures are similar -- they have line-of-sight velocities consistent with prograde motion in circular orbits at $v_{GSR} = 236~\kms$, shown as the dashed curves in the bottom panel of Figure~\ref{fig:compare}. For the M giants with S/N $>$ 25, the heliocentric distances of individual stars in each population (calculated in the same way as described in Section~\ref{sec:distance}) are presented in the middle panel of Figure~\ref{fig:compare}. Although the uncertainty in distance measurements is large ($\sim20\%$), there is a clear trend that the A13 stars with $l > 180 \degree$ tend to be at similar distances to GASS/Mon while those with $l < 180 \degree$ tend to be at larger distances, more like the stars in TriAnd. 

As described in Section~\ref{sec:feh}, we derived the metallicity of A13 stars using the CaT index. We developed an empirical calibration relation in a similar way to TriAnd in~\citet{Sheffield2014} to minimize any systematic effects in metallicities and distances across different samples. For GASS/Mon, \citet{Crane2003} used two of the three CaT lines  ($\lambda\lambda$8498, 8542) and one Mg \textsc{i} ($\lambda\lambda$8807) line as the spectral indices. The derived mean \feh is $-$0.4 $\pm$ 0.3 for GASS/Mon, which is slightly more metal-rich than what we get for A13, i.e., \feh = $-$0.57 $\pm$ 0.21. 
Meanwhile, \citet{Sheffield2014} derived the mean \feh from the CaT index of \feh = $-$0.62 $\pm$ 0.44 for TriAnd1 and \feh = $-$0.63 $\pm$ 0.29 for TriAnd2, which are very close to the metallicity derived for A13\footnote{Note the quoted uncertainties for A13, GASS, TriAnd are all calculated as the standard deviation of the metallicities of individual stars.}. \citet{Sheffield2014} also derived the metallicity by fitting a grid of isochrones to 2MASS and SDSS photometry data simultaneously. The metallicity derived from isochrone fitting (\feh $\sim$ $-$0.9) tends to be more metal-poor than the metallicities derived from the CaT index. Such a systematic bias might also present in A13 and GASS/Mon. Metallicity measurements from high-resolution spectroscopy is needed to provide accurate [Fe/H] and will be presented in a future paper (Sesar et al. in prep). 

\subsection{Comparison with the Pan-STARRS Substructures Map}\label{sec:panstarr}

\begin{figure*}[th!]
\centering
\epsscale{0.8}
\plotone{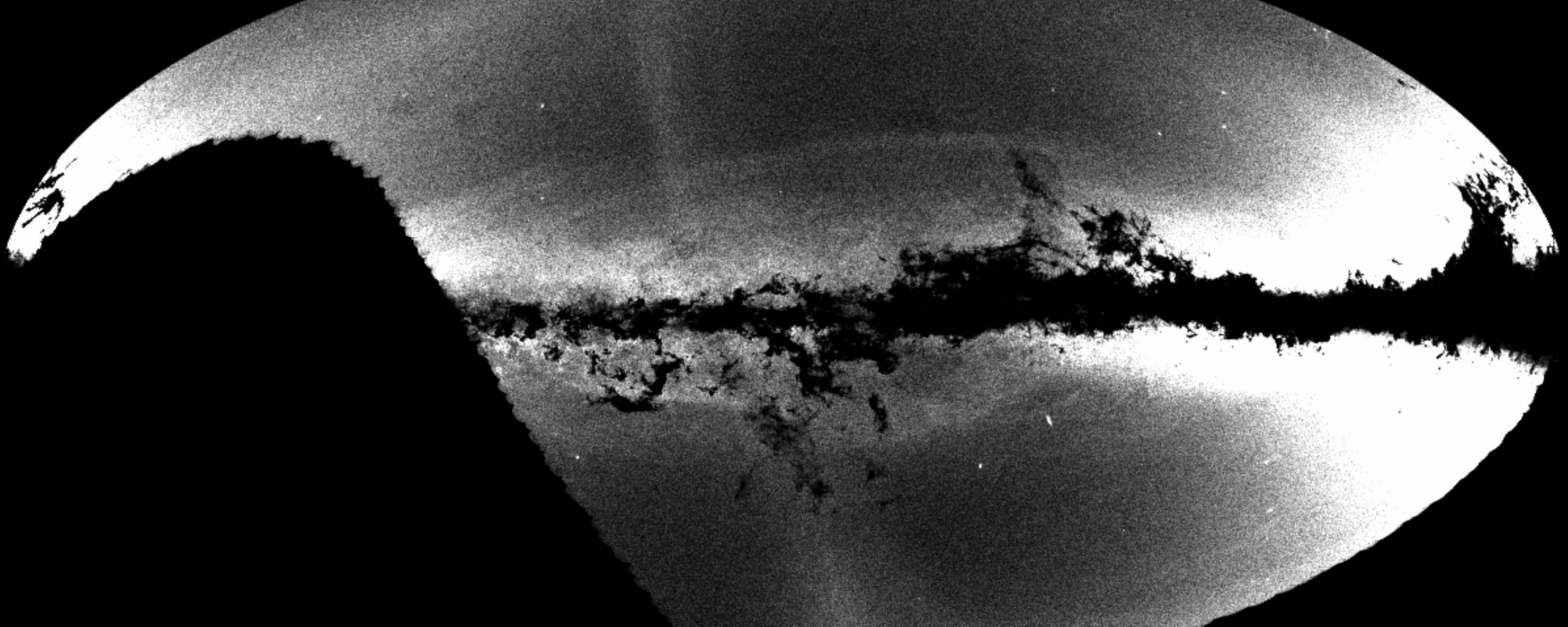}
\plotone{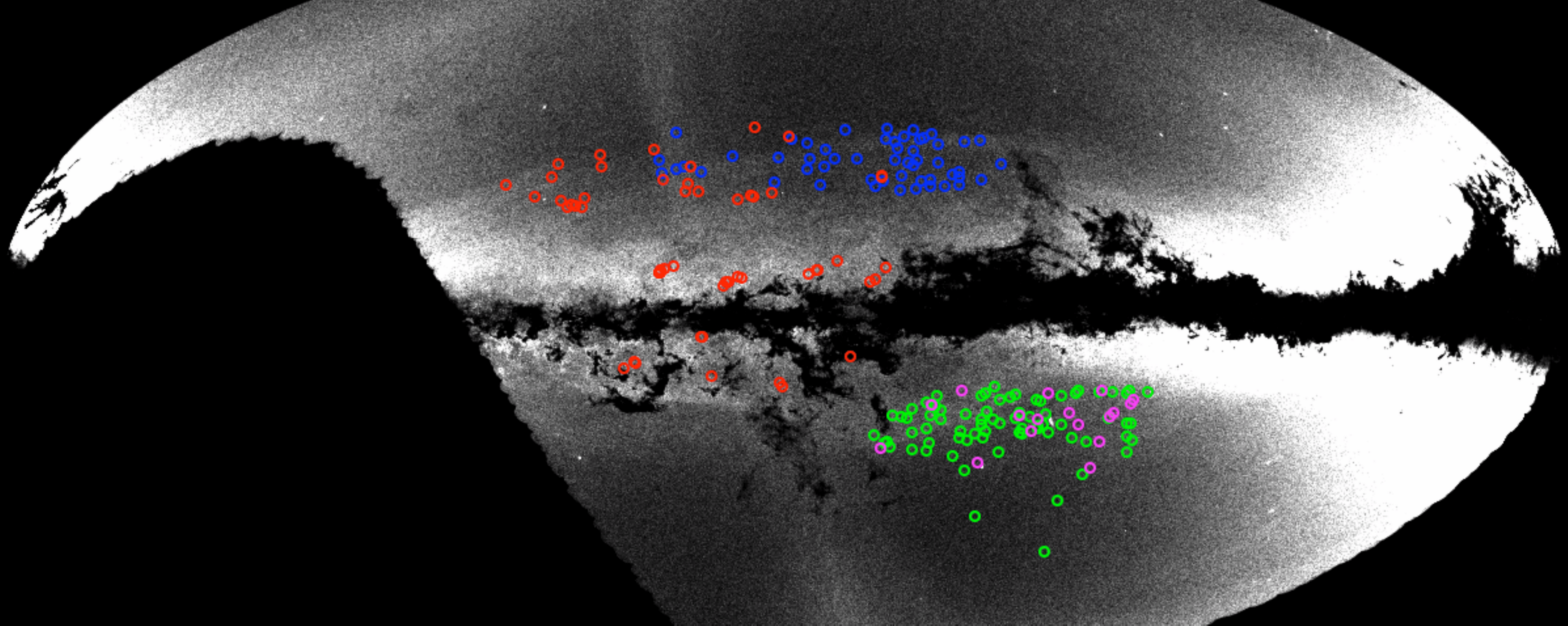}
\caption{(Top) Stellar density map (in Galactic coordinates) of main-sequence turnoff stars at $\sim 16$~kpc from Pan-STARRS catalog, with brighter areas indicating higher surface densities. The Galactic anticenter is in the middle. (Bottom) Same density map but overplotted with M giants from GASS (red), A13 (blue), TriAnd1 (green), TriAnd2 (magenta). The M giants well-trace the overdensities seen in Pan-STARRS, especially for the A13 stars. The vertical structure in the density map is the Sagittarius tidal stream.}
\label{fig:panstarrs}
\end{figure*}

We further compare these structures with the main-sequence turnoff star density map from the work of \citet{Bernard2016} based on the Pan-STARRS catalog~\citep{panstarr}.
Figure~\ref{fig:panstarrs} is similar to their Figure 1 but in Galactic coordinates and at a heliocentric distance of $\sim16$~kpc, which is close to the average distance of A13. The M giants, especially in A13, is in good positional alignment with the structures in the Pan-STARRS map. This coincidence further suggests that A13 is likely to be an extension of GASS towards lower Galactic longitude (and further distance) as discussed in Section~\ref{sec:gass}. Southern overdensity seems to match the TriAnd structure in the density map, but less prominently. 
The clear vertical structure in the density map is the Sagittarius tidal stream. We therefore expect some contamination of non-member stars from Sagittarius stream which may slightly inflate the velocity dispersion of A13, as mentioned in Section~\ref{sec:rv}.

\citet{Slater2014} also made similar density maps with main-sequence turnoff stars using an earlier version of the Pan-STARRS catalog. It is worth-noting that our M giant samples are in good agreement with the Monoceros Ring features highlighted in their paper (see, e.g., Features A, B, C and D in the Figure 3 of~\citet{Slater2014}).

\subsection{Towards a Unified Picture?}
Overall, our study suggests similar kinematical and spatial properties and trends with position in the Galaxy for all three major anticenter stellar structures -- TriAnd (including TriAnd1 and TriAnd2), GASS/Mon and A13 -- at least when using M giants as the stellar tracer. These commonalities are strongly suggestive of physical connections, and even possibly a single origin for all of these features. Hypotheses for the origin of these structures span from them being associated with the Galactic disk to having been accreted from an infalling satellite galaxy.

\citet{Xu2015} hypothesized that overdensities like GASS/Mon and TriAnd could be the large-scale signatures of vertical oscillations of the Galactic disk. 
Assymmetries in the velocity and spatial distributions of stars above and below the plane of the Galaxy in the vicinity of the Sun had already been detected both in SDSS \citep{Widrow2012} and the RAVE survey \citep{Williams2013}.
\citet{Xu2015} proposed that  GASS/Mon (closer to the Sun) and TriAnd (farther from the Sun) could be associated with the same locally-apparent disturbance, as the northern and southern parts of a vertically oscillating ring propagating outward from the Galactic center. 
\citet{PriceWhelan2015} provided the first concrete support for this hypothesis with evidence that the Galactic disk was the original birth-place of stars in TriAnd. They found a very low number ratio of RR Lyrae to M giant stars in TriAnd, consistent with the metal-rich stellar population of the disk and quite unlike those seen in surviving satellite galaxies. Many recent simulation studies have tried to understand the origin of such vertical structures of the Milky Way disc~\citep[e.g.][]{Gomez2017}.
Using N-body and/or hydrodynamical simulations, \citet{Laporte2016} and \citet{Gomez2016} have shown that Milky Way satellites could produce strong disturbances and might lead to the formation of vertical structure in the Galactic disk.

\begin{figure}[th!]
\centering
\epsscale{1.2}
\plotone{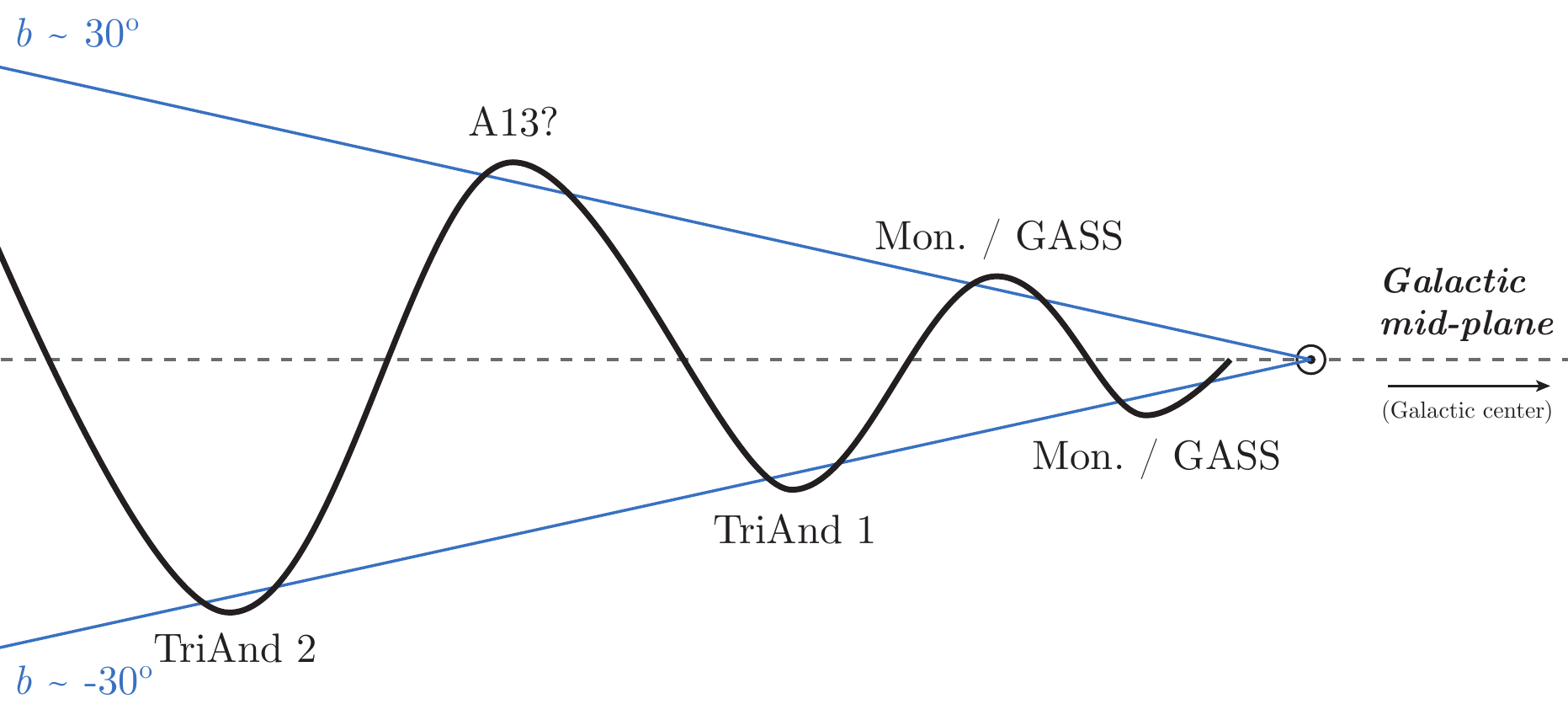}
\caption{Illustration of a possible scenario where GASS, A13, TriAnd1, and TriAnd2 are the results of ringing disk oscillations. In this scenario, GASS and A13 are two sequences of the northern rings, while TriAnd1 and TriAnd2 are the two sequences of the southern rings. Symbol $\odot$ indicates the location of the Sun.}
\label{fig:cartoon}
\end{figure}

\begin{figure*}[th!]
\centering
\epsscale{0.8}
\plotone{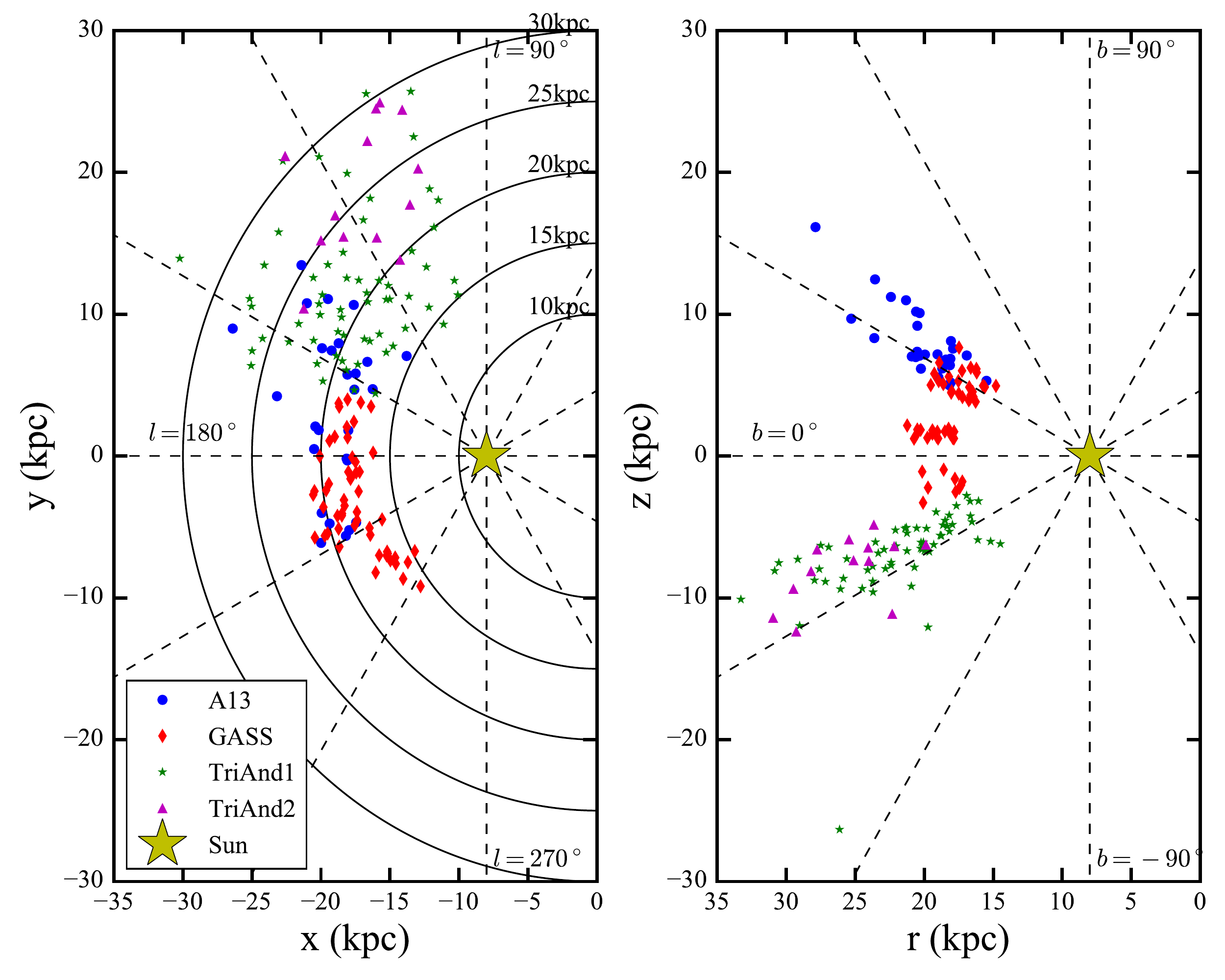}
\caption{Distributions of M giants in A13, GASS, TriAnd1 and TriAnd2 shown in Galactocentric Cartesian coordinates in the $x-y$ plane (left) and in the $r-z$ plane (right), where $r = \sqrt{x^2+y^2}$. The Galactic Center is at (0, 0, 0) and the Sun is at ($-$8, 0, 0)~kpc. Galactic longitude and latitude (dashed) and curves at constant Galactocentric radius (solid) are shown.}
\label{fig:cartesian}
\end{figure*}

Figure~\ref{fig:cartoon} schematically illustrates a possible scenario where these overdensities are the signatures of the disk oscillations and have different Galactocentric distances: in this scenario, GASS/Mon and A13 would be two sequences of northern rings while TriAnd1 and TriAnd2 would be two sequences of southern ring-like structures.
We note, however, that this cartoon is an over-simplification of the complex, more spiral-like structures tha can be formed from disk perturbations (see, e.g., simulations of disk fly-by's and Figure~8 in \citealt{PriceWhelan2015}).
Figure~\ref{fig:cartesian} explores this hypothesis further by plotting projections of the three-dimensional distributions of M giants in all these structures, using the approximate heliocentric distances derived from photometry and CaT metallicity. 
As the distance measurements have very large uncertainty, it is hard to determine from these projections whether the description of these overdensities as concentric rings or wrapped spiral structure in \cite{Xu2015} and \cite{PriceWhelan2015} is really an accurate representation of their morphology.
It is also worth noting that the large overlaps in distances of the GASS and A13 stars, as well as TriAnd1 and TriAnd2 stars, as shown in Figure~\ref{fig:cartesian}, do not support the picture in Figure~\ref{fig:cartoon} where a clear gap in distance should exist between the two sequences. This could be explained  (1)  either by the large distance uncertainties in the data, which blur the gap between the two sequences (as~\citealt{Martin2007} show with distinct distances for TriAnd1 and TriAnd2 for turnoff stars using MegaCam data), or (2) by a more complicated disk oscillation model where the multiple sequences may have overlap in distance depending on the azimuthal line-of-sight.

From the evidence above, we do not yet have a clear enough picture to conclusively prove that stars in these overdensities were born in the disk rather than in an accreted satellite galaxy.
Morphologically, concentric rings and/or arcing overdensities can be produced in either the ringing disk \citep[see, e.g.,][]{Gomez2013,PriceWhelan2015} or satellite accretion model \citep[see, e.g.,][]{Penarrubia2005,Sheffield2014}. \citet{Slater2014} also compare the Pan-STARRS density maps with mock data from simulations and show that those stream-like features can be produced by either tidal debris of a dwarf galaxy or large disk distortion.  
While the stellar populations of TriAnd have been show to be more like the disk than of known satellites of the Milky Way \citep{PriceWhelan2015}, detailed chemical abundance pattern analyses for stars in GASS/Mon and TriAnd show that these structures are more likely to be reminiscent of satellite galaxies~\citep{Chou2010,Chou2011}.  Furthermore, \citet{Chou2011} also show that the chemical abundance patterns of TriAnd are distinct from those of GASS/Mon, suggesting the two systems are unrelated. A more extensive and complete study of the abundance patterns and stellar populations of A13 and a comparison with GASS/Mon and TriAnd will help distinguish one scenario from the other.

Proper motions could also play a deciding role: \citet{Sheffield2014} found that the magnitude difference of (i.e., spatial offsets between) TriAnd1 and TriAnd2 could only be produced by a satellite disrupting on a retrograde orbit with respect to the disk, while kicked-out disk material would be expected to be on prograde orbits. As these M giants are relatively bright (V$\sim$13--16~mag), these hypotheses could be tested with the proper motions from the upcoming $Gaia$ data release.

Admittedly, as the M giant samples discussed here for GASS, A13 and TriAnd have different selection criteria (e.g., magnitude and color selection, Galactic extinction masks, etc.) between different studies, our comparisons might be affected by selection effects. Therefore, a consistent target selection and analysis of the M giants at low Galactic latitude with ongoing or future surveys (e.g., LAMOST, APOGEE, DESI, etc) will provide a better understanding of these structures near the anticenter of the Galaxy.

\section{Conclusions}\label{sec:conclusion}

This paper presents a study of A13, an overdensity of M giant star counts reported in \citet{Sharma2010}. We derived the kinematics and metallicities of candidate members of A13 via moderate-resolution spectroscopic observations. Our results support two key conclusions.

First, the candidate M giant members have a relatively small velocity dispersion ($\lesssim$ 40 \kms), implying that A13 is a genuine structure rather than the chance super-position of random halo stars. The confirmation of the A13 structure is an interesting result by itself, as it demonstrates the ability of the \citet{Sharma2010} group finding algorithm to find substructures in large-scale photometric stellar catalogs.

Second, from the position of A13 on the sky and its kinematic properties, A13 may be associated with two other known substructures in this region: the GASS/Mon and TriAnd overdensities. The radial velocity of the stars in A13 follow the same trend in Galactic longitude as the stars in both GASS/Mon and TriAnd and the stars in each have similar dispersions. 

The data collected so far --- with large errors on metallicities and distances --- do not allow us to map the morphology and motions of these structures with enough resolution to present a conclusive single scenario for the nature of these overdensities. Further studies on chemical abundance and stellar populations are necessary to understand the nature of these structures. In particular, a more extensive and complete spectroscopic analysis of these three structures with ongoing or future surveys (e.g. LAMOST, APOGEE, DESI) will provide a better understanding of these structures near the anticenter of the Galaxy.

\acknowledgements
The authors thank Jeff Crane for sharing the C03 data and for helpful conversations. TSL acknowledges support by the Mitchell Institute Fellowship from Texas A\&M University. KVJ, APW and AAS contributions were partially funded by National Science Foundation grant AST-1312186. SRM, GJD and WR contributions were partially funded by National Science Foundation grant AST-1312863. The authors thank the anonymous referee for suggestions that improved the quality of the paper. The authors also thanks the useful conversations with Brian Yanny and Jeff Carlin.

\bibliographystyle{apj}
\bibliography{bib}

\end{document}